\documentclass[%
aps,
jmp,%
amsmath,amssymb,
reprint,%
]{revtex4-1}
\usepackage{graphicx}% Include figure files
\usepackage{multirow}
\usepackage{diagbox}
\usepackage{subfigure}
\usepackage{dcolumn}% Align table columns on decimal point
\usepackage{bm}% bold math
\usepackage[mathlines]{lineno}% Enable numbering of text and display math
\usepackage{epstopdf}
\usepackage{amsmath}
\usepackage[colorlinks]{hyperref}
\hypersetup{
	colorlinks = true,
	citecolor = {blue},
	linkcolor = {red}
}

\begin{document}
	\title{Experimental secure entanglement-free quantum remote sensing over 50 km of optical fiber  }% Force line breaks with \\
\author{ Wenjie He,$^{1}$ Chunfeng Huang,$^{1}$ Rui Guan,$^{1}$ Ye Chen,$^{1}$ Zhenrong Zhang,$^{2}$ Kejin Wei$^{1,*}$ }

\address{
	$^1$Guangxi Key Laboratory for Relativistic Astrophysics, School of Physical Science and Technology,
	Guangxi University, Nanning 530004, China\\		
	$^1$Guangxi Key Laboratory of Multimedia Communications and Network Technology, School of Computer Electronics and Information, Guangxi University, Nanning 530004, China\\		
	$^*$kjwei@gxu.edu.cn
}
\date{\today}

\begin{abstract}

	Secure quantum remote sensing (SQRS) uses quantum states to gather information about distant objects or environments while ensuring secure data transmission against eavesdropping. It has potential applications in various fields, including environmental monitoring, military surveillance, and disaster response, where both data accuracy and transmission security are critical. Recent experiments have demonstrated the feasibility of SQRS using entanglement states. Here, we experimentally demonstrate an SQRS that can estimate a phase without requiring entanglement, offering the practical advantage that single-qubit states are easier to prepare. We successfully estimate the preset phase information at a remote site over a fiber distance of  50 km, which serves as a key step toward long-distance applications.

\end{abstract}

\maketitle

\section{introduction}

Quantum information science, which incorporates quantum properties into information processing tasks, offers unique advantages compared to its classical counterparts. For instance, quantum cryptography, such as quantum key distribution (QKD)~\cite{1992-Bennett,QKD-2009-Scarani,2014-BENNETT,2014-Lo,QKD-2019-Pirandola,2020-xu,2021-Kwek,QKD-2023-Zapatero}, can enable information-theoretic secure communications over long distances~\cite{QKDLD-2015-Korzh,QKDLD-2018-Boaron,QKDLD-2021-Pittaluga,QKDLD-2022-Wang,QKDLD-2023-Zhou,QKDLD-2023-liu}. Furthermore, quantum metrology has demonstrated the capacity to surpass the standard quantum limit that restricts the sensitivity of established classical methods, and, in certain scenarios, it can even achieve the Heisenberg limit~\cite{2004-Giovannetti,2006-Giovannetti,2007-Nagata,2011-Giovannetti,2014-Toth,2018-Chen-1,2018-Chen,2018-Zhou,2020-Emanuele,2021-Liu,2024-Hotter}.

The primary focus of quantum information science lies in the areas of computational power~\cite{QC-2019-Arute,QC-2021-ARRAZOLA,QC-2024-Bluvstein}, communication security~\cite{QSS-1999-Hillery,QDS-2001-Gottesman,QSS-2013-Wei,QDS-2021-Qi,QDS-2024-Cao,QDS-2024-Du,QSS-2024-Xiao,QSS-2024-Wang}, and measurement precision~\cite{MP-2024-Hayashi,MP-2024-Hotter,MP-2024-Yang}. Numerous studies have revealed the interconnected nature of these three domains, showing that their intersection fosters innovative approaches in quantum information processing. For instance, the combination of quantum computation and quantum cryptography gives rise to blind quantum computing~\cite{BQC-2012,BQC-2012MORMAE,BQC-2013-MANTRI,BQC-2018-SHENG}, allowing a client with limited computational resources to delegate universal quantum computations to a remote server equipped with a powerful quantum computer, while ensuring that the server cannot access the computational information~\cite{BQC-2012Barz,BQC-2019-Jiang,BQC-2023-POLACCHI,2021-Huang}.

Secure quantum remote sensing (SQRS) is a recent emerging technology~\cite{2019-Takeuchi} that combines the benefits of quantum metrology and quantum communication~\cite{2019-Huang} to achieve secure and precise estimation of unknown parameters at remote sites. The security of SQRS has been demonstrated through its secrecy capacity for information bits and asymmetric Fisher information gain for sensing. Several theoretical advances have been made, including the extension of applications~\cite{2024-lIU,2024-Moore}, enhancement of security~\cite{2022-Peng}, and improvement of efficiency~\cite{2022-sHETTELL}. In experiments, the feasibility of SQRS has been demonstrated using entangled sources~\cite{2020-Yin}. However, the preparation of entangled states still faces technical challenges, which could compromise the practical applications of SQRS.

In this study, we experimentally demonstrate an entanglement-free SQRS protocol~\cite{2023-moore} that protects sensing information without using entangled states.  To address the challenges introduced by practical imperfections, we develop a pre-calibration technique, which enabled the protocol to efficiently estimate the phase in a real-world setup. We successfully estimate the preset phase information with high precision  over 50 km of optical fiber. The results demonstrate the feasibility of SQRS without entanglement over long distances and serve as a key step toward its practical implementation.

The organization of this paper is outlined as follows.  In Section~\ref{protocol}, we introduce the employed entanglement-free SQRS protocol and develop a pre-calibration technique. In Section~\ref{setup}, we provide a detailed description of the SQRS experimental setup. In Section~\ref{Experiment and result}, we verify the protocols in a real-world system. Finally, we provide the conclusion in Section~\ref{conclusion}.
	\section{Entanglement-free SQRS protocol}\label{protocol}

	    \begin{figure}
		\centering 
		\includegraphics[width=1\linewidth]{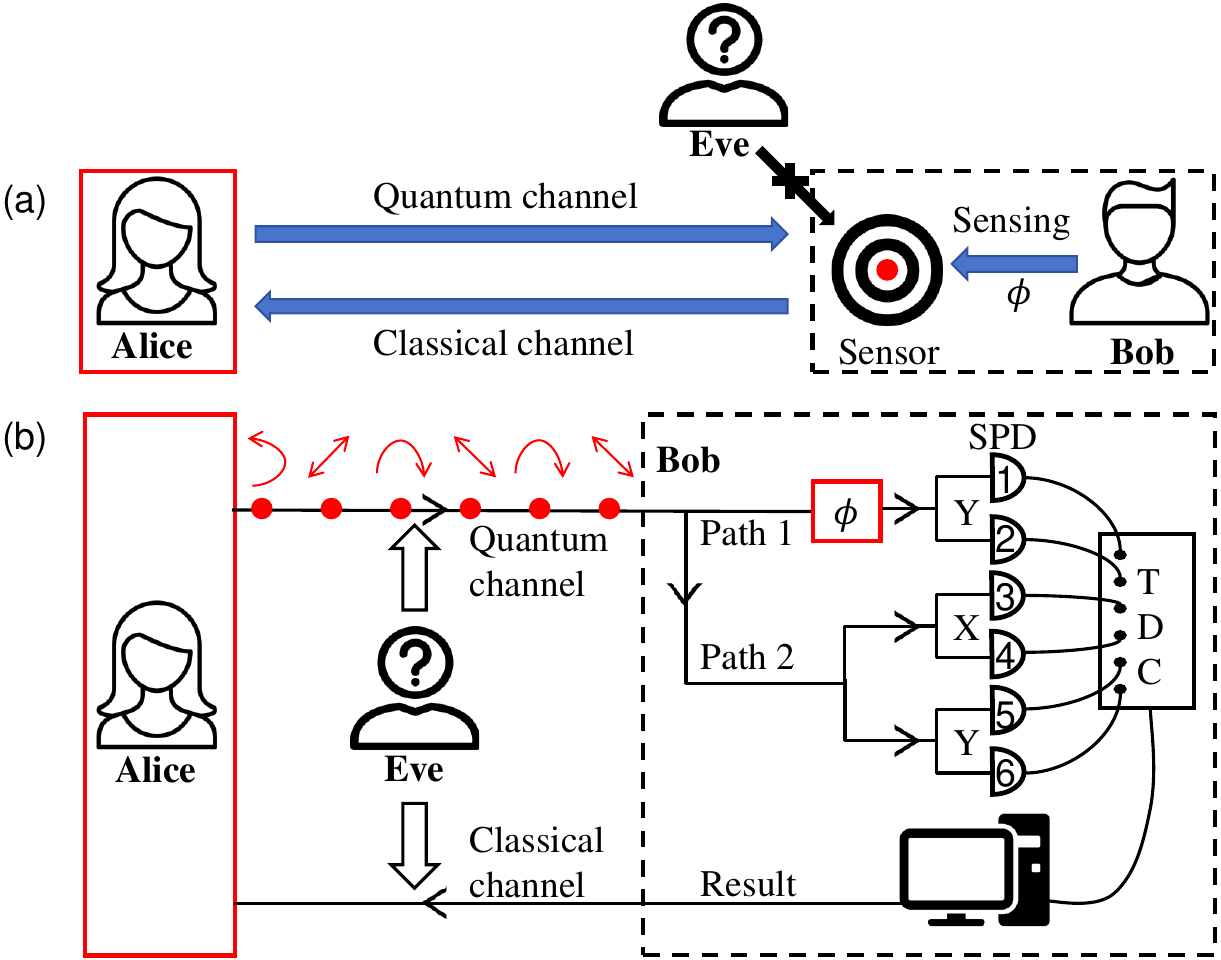}
		\caption{(a) A typical application scenario of SQRS. (b) The SQRS protocol whitout entanglement: Alice prepares the four polarization states and sends them to Bob. Upon arrival at Bob's end, the photons are randomly split into two beams, one of which is used to sense the unknown phase, while the other is used to ensure the security of the quantum channel. The measurement results and their corresponding detectors are then sent to Alice through a classical channel. Eve can attack both the quantum and classical channels in an attempt to obtain information about $\phi$. In the figure, the red box indicates the security zone, and the black dotted box indicates the non-security zone.}\label{fig1}
	\end{figure}

   In this section, we briefly review the SQRS scenario and the original entanglement-free protocol, which was first proposed in Ref.~\cite{2023-moore}, in an ideal case. Furthermore, we develop a pre-calibration technique that enables the protocol to efficiently estimate the phase in a practical setup with dark counts and optical misalignment.
   
   \subsection{Original protocol}
    A typical scenario is illustrated in Fig.~\ref{fig1}(a): a doctor (Alice), located within a secure area, wishes to remotely estimate the phase of a sensor carried by a patient (Bob) outside this secure area to assess the patient's health status. To protect the patient's privacy, Alice aims to prevent any potential eavesdropper (Eve) from obtaining the phase. Although Bob is not in a secure node, he is a trusted participant committed to maintaining the confidentiality of his health status. Therefore, Bob follows Alice's instructions and takes the necessary actions to prevent Eve from directly accessing the phase in the health sensor.
   
   Our entanglement-free SQRS protocol achieves secure quantum remote sensing by transmitting single-photon qubits, as illustrated in Fig.~\ref{fig1}(b). The detailed steps are as follows:

\textbf{Step 1: State preparation and transmission.} Alice and Bob share a quantum channel and a classical channel, both of which may be vulnerable to attacks by Eve. Alice prepares a sequence of quantum states $|\psi_{\text{in}}\rangle$, where each state is randomly chosen from the eigenstates of $\sigma_x$ and $\sigma_y$ and can be expressed as $|\psi_{\text{in}}\rangle \in \left\{ \frac{1}{\sqrt{2}} (|H\rangle \pm |V\rangle), \frac{1}{\sqrt{2}} (|H\rangle \pm i|V\rangle) \right\}$. The prepared states are transmitted to Bob via the insecure quantum channel. Note that only Alice knows the specific quantum states being transmitted.

\textbf{Step 2: Phase introduction and  measurement.} Upon reaching Bob, the quantum states are randomly directed along one of two paths. In the first path (denoted as Path 1), the quantum state interacts with the phase ($\phi$) in the sensor, generating an encoded state $|\psi_{\phi}\rangle$, which is expressed as $\frac{1}{\sqrt{2}} (|H\rangle + e^{i\phi}|V\rangle)$ when $|\psi_{\text{in}}\rangle = \frac{1}{\sqrt{2}} (|H\rangle + |V\rangle)$. The encoded states are then measured using single-photon detectors (SPD-1 and SPD-2) in the $\sigma_y$ basis. In the second path (denoted as Path 2), the quantum state is randomly measured in a basis chosen from $\sigma_x$ and $\sigma_y$ and then detected using SPD-3 to SPD-6. In fact, the process in the second path follows the typical BB84 quantum key distribution protocol~\cite{2014-BENNETT}, which enables the sharing of secure key bits with information-theoretic security. Bob record the which detector click and corresponding time slot, and then sends these information  to Alice via the classical channel. In principle, Eve can obtain these information since the classical channel is assumed to insecure.

\textbf{Step 3: Phase  estimation.} Upon receiving the measurement results, Alice performs Bayesian analysis to estimate the phase from the data in Path 1 and evaluates the fidelity of the quantum states received by Bob using the data from Path 2. Specifically, in Path 1, the four polarization states ($|\psi_{\text{in}}\rangle$) sent by Alice evolve into encoded states ($|\psi_{\phi}\rangle$) after interacting with the phase ($\phi$). When these states are projected onto the $\sigma_y$ basis, these states collapse into one of two possible outcomes: $|\psi_{\text{out}}\rangle \in \left\{ \frac{1}{\sqrt{2}} (|H\rangle + i|V\rangle), \frac{1}{\sqrt{2}} (|H\rangle - i|V\rangle) \right\}$. Under ideal conditions, the probability of observing each outcome is determined by $p_{i}(out|\phi) = |\langle \psi_{\text{out}} | \psi_{\phi} \rangle|^2$, where $i \in {1, 2, \dots, 8}$. Table~\ref{table1} shows the measurement probabilities $p_{i}^\phi$ when Alice sends $\sigma_x$ and $\sigma_y$ eigenstates to Bob, who encodes $\phi$ on them and measures in the $\sigma_y$ basis. 

Suppose that in an experiment, two SPDs in path 1 detect a total of $m$ events. Alice records, based on Alice's prepared states and which detector clicks, the numbers of each measurement outcome listed in Table~\ref{table1} $n_i^\phi$, where $i \in \{1, 2, \dots, 8\}$, and $m = \sum^8_{i=1} n_i^\phi$. 
To estimate the phase, Alice constructs a likelihood function, which can be expressed as 

\begin{multline} \label{eq1}
	L(\varphi) \propto (1+\sin{\varphi})^{n_1^\phi+n_4^\phi}(1-\sin{\varphi})^{n_2^\phi+n_3^\phi}\\
	\times(1+\cos{\varphi})^{n_5^\phi+n_8^\phi}(1-\cos{\varphi})^{n_6^\phi+n_7^\phi}.
\end{multline}

The estimated phase $\hat{\phi}$ amount to the phase at which the function's value is maximized, which is denoted as
\begin{equation} \label{eq2}
	\hat{\phi}= \arg\max_{\varphi \in [0, 2\pi]} L(\varphi).
\end{equation}

    \begin{table}[htbp]
    	\centering
    	\setlength{\tabcolsep}{8pt}
    	\renewcommand{\arraystretch}{1.5}
    	\caption{The measurement probabilities $p_{i}$ when Alice sends the four states to Bob, who encodes phase ($\phi$) on them and measures in the $\sigma_y$ basis.}
    	\label{table1}
    	\begin{tabular}{|c|c|c|}
    		\hline
    		\multirow{2}{*}{\textbf{Eigenstate}} & \multicolumn{2}{c|}{\textbf{Measurement}} \\ \cline{2-3} 
    		& \textbf{$\sigma_y=0$} & \textbf{$\sigma_y=1$} \\ \hline
    		$\sigma_x=0$ & $p_{1}^\phi=\frac{1}{2}(1+\sin{\phi})$ &$p_{2}^\phi=\frac{1}{2}(1-\sin{\phi})$          \\ \hline
    		$\sigma_x=1$ & $p_{3}^\phi=\frac{1}{2}(1-\sin{\phi})$ &$p_{4}^\phi=\frac{1}{2}(1+\sin{\phi})$          \\ \hline
    		$\sigma_y=0$ & $p_{5}^\phi=\frac{1}{2}(1+\cos{\phi})$ & $p_{6}^\phi=\frac{1}{2}(1-\cos{\phi})$          \\ \hline
    		$\sigma_y=1$ & $p_{7}^\phi=\frac{1}{2}(1-\cos{\phi})$ & $p_{8}^\phi=\frac{1}{2}(1+\cos{\phi})$          \\ \hline
    	\end{tabular}
    \end{table}

\subsection{Pre-calibration technique}

\begin{figure}
	\centering
	\hspace{-1cm} 
	\includegraphics[width=1\linewidth]{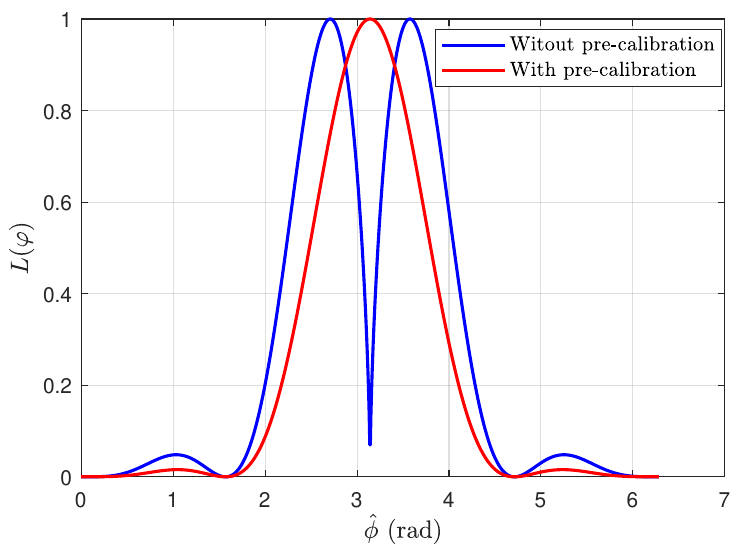}
	\caption{The relationship between the likelihood function ($L(\varphi)$) and the phase estimate ($\hat{\phi}$) when estimating $\pi$. The blue (red) line represents the relationship between $L(\varphi)$ and $\hat{\phi}$ without (with) using pre-calibration technique.}\label{fig2}
\end{figure}

The aforementioned protocol can efficiently estimate the phase in an ideal case where the system has no dark counts or optical misalignment. However, this case cannot be fulfilled in a practical setup. Specifically, as shown by the blue line in Fig.~\ref{fig2}, when estimating the phase $\phi=\pi$, the likelihood function has two maximum values due to non-zero $n_5^\pi$ and $n_8^\pi$ (which should be zero in an ideal case), resulting in a failed phase estimation. 

To address this issue, we develop a pre-calibration technique. In this method, Alice sends $m_\theta$ pulses and asks Bob to randomly encode a pre-set phase $\theta\in\{0,\pi/2,\pi,3\pi/2\}$. Then Bob provides feedback involving the encoded phase, which detector clicked, and the corresponding time slot via the classical channel. Alice can estimate eight non-zero probabilities, denoted as $\{p_1^{3\pi/2},p_2^{\pi/2},p_3^{\pi/2},p_4^{3\pi/2},p_5^{\pi},p_6^0,p_7^0,p_8^{\pi}\}$, where the subscript denotes the encoded phase. Finally, the probabilities of observing each outcome are revised as
\begin{equation}
	p_{i}^{\phi\prime} = p_{i}^{\phi}  -  p_{i}^{\theta}, \quad i\in\{1,2,...,8\}.
\end{equation}
Using the revised probabilities, the likelihood function only has one maximum value (as the red line in Fig.~\ref{fig2}) which enables estimate the correct phase with high precision.

It is important to emphasize that this method is applied only during the post-processing stage, and does not introduce any security vulnerabilities.

	\section{SETUP}\label{setup}
	
    \begin{figure*}
	\centering
	\includegraphics[width=0.7 \textwidth]{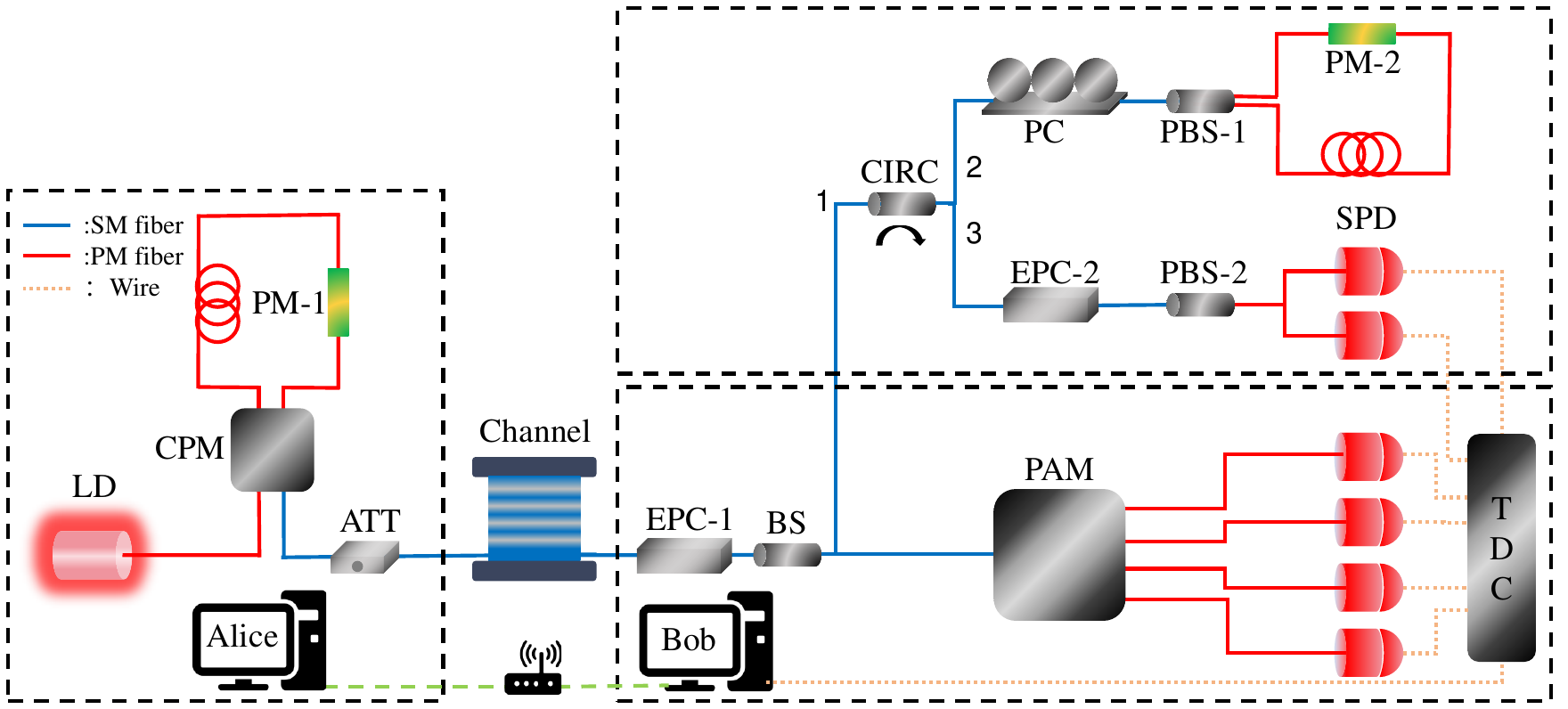}
	\caption{Schematic diagram of experimental device. LD: 1550 nm commercial laser source; CPM: customized polarization module; PM: phase modulator; ATT: attenuator; EPC: electronic polarization controller; PAM: polarization analysis module; SPD: single-photon avalanche detector; TDC: time-to-digital converter; PM fiber, polarization maintaining fiber; SM fiber, single mode fiber; Wire: electric wire. 
	}\label{fig3}
\end{figure*}

	We demonstrate the entanglement-free SQRS using a fiber-optic setup. The experimental setup is shown in Fig.~\ref{fig3}. At Alice's site, a weak coherent source, generated using a phase-randomized laser (LD) and an attenuator (ATT), is used to probabilistically generate single-photon states. The eigenstates of $\sigma_x$ and $\sigma_y$ are generated using a Sagnac-based polarization modulation module~\cite{2019-Li,2021-Ma}. The module consists of a customized polarization beam splitter (CPM) and a phase modulator (PM-1) within the Sagnac interferometer. The CPM features a polarized main fiber input aligned at a 45° angle, splitting optical pulses into two orthogonally aligned linear polarizations. Pre-calibrated voltages are applied to PM-1 using an arbitrary waveform generator (AWG) to generate one of the eigenstates, denoted as $|\psi_{\text{in}}\rangle \in \left\{ \frac{1}{\sqrt{2}} (|H\rangle \pm |V\rangle), \frac{1}{\sqrt{2}} (|H\rangle \pm i|V\rangle) \right\}$, which is then sent to Bob through commercial optical fiber.
	
	At Bob's side, an electronic polarization controller (EPC-1) compensates for polarization drift during fiber transmission. The light pulse is then directed along one of two paths using a beam splitter (BS). In the first path, the light pulse enters port 1 of the circulator, exits from port 2, and then enters a phase modulation module based on a Sagnac interferometer. The Sagnac interferometer consists of a polarization controller (PC), a polarization beam splitter (PBS-1), and a phase modulator (PM-2). The PC compensates for polarization drift during the BS and PBS-1. PBS-1 splits the pulse into two orthogonally polarized pulses. By triggering the PM-2, a phase $\phi$ is encoded into one of the polarized pulses, generating the phase-introduced state $|\psi_{\phi}\rangle$. The light pulse is then measured in the $\sigma_y$ basis, implemented using EPC-2 and PBS-2, and the photons are detected by two gated single-photon detectors (SPDs) with an average detection efficiency of 4.5$\%$.
	
	In the second path, the light pulse enters the polarization analysis module (PAM)~\cite{2021-Ma}, which measures the $\sigma_x$ and $\sigma_y$ bases. It includes a 50/50 BS and two polarization-maintaining PBSs. The measurement bases for BB84 are aligned with the module, with one of the input PMFs of the PBSs vertically aligned with the BS, and the other rotated at an angle of 45$^\circ$. The photons are detected by four SPDs. All detection events are recorded using a time-to-digital converter (TDC). The recorded data are processed using a personal computer and sent to Alice via Wi-Fi.

	\section{Results}\label{Experiment and result}
	
	To validate the performance of the entanglement-free SQRS protocol, we conduct several tests using the previously described setup over a 50-km optical fiber.
	
	First, we check the quantum bit error rate (QBER) of the setup, estimate from the error counts of the four detectors in Path 2. The obtained QBER is lower than 6\%, which meets the SQRS requirement~\cite{2023-moore}. Then, Alice send approximately $2.1 \times 10^4$ pulses to estimate the pre-calibration parameters and obtained the following results: $p_1^{3\pi/2} = 0.0229$, $p_2^{\pi/2} = 0.0447$, $p_3^{\pi/2} = 0.0394$, $p_4^{3\pi/2} = 0.0161$, $p_5^{\pi} = 0.0732$, $p_6^0 = 0.0164$, $p_7^0 = 0.0118$, and $p_8^{\pi} = 0.067$.

\begin{figure}
	\centering 
	\includegraphics[width=1\linewidth]{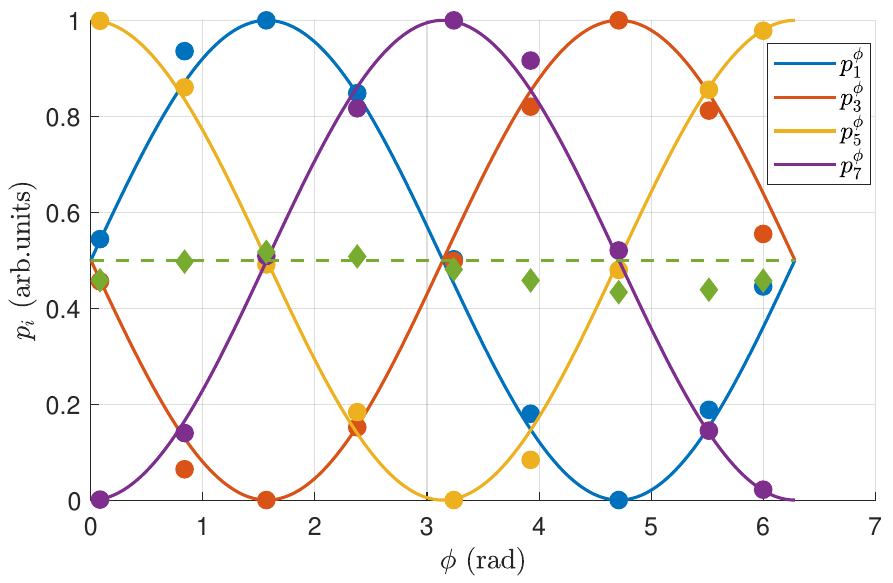}
	\caption{ Dependence of probabilities on $\phi$. The blue, orange, yellow, and purple curves represent the response probabilities of SPD-1 in the ideal case, when Alice sends four different polarization states ($\sigma_x = 0$, $\sigma_x = 1$, $\sigma_y = 0$, $\sigma_y = 1$). The blue, orange, yellow, and purple dots correspond to the experimental response results for these states. The green dashed line represents the information that both Bob and a third party can obtain, specifically the ratio between the counts of SPD-1 and the total counts from both SPD-1 and SPD-2, with an ideal value of 0.5. The green diamonds represent the experimental values of this ratio.
	}\label{fig4}
\end{figure}

Next, we measure nine phases, denoted as $\phi_k$ ($k = 1, 2, \dots, 9$), ranging from 0 to $2\pi$. The obtained probabilities for SPD-1 are shown in Fig.~\ref{fig4} for the four polarization states. It can be seen that the obtained probabilities  agree with the theoretical predictions. Furthermore, we evaluate the ratio between the SPD-1 counts and the total counts from SPD-1 and SPD-2, which can be expressed as $(n_1+n_3+n_5+n_7)/\sum_{i=1}^8 n_i$. Since this ratio is the only information Eve can steal for phase estimation, the obtained values for each phase are plotted in Fig.~\ref{fig4}. It can be seen that the ratio remains almost constant as $\phi$ varies, indicating that Eve cannot extract meaningful information from the encrypted data.

\begin{figure}
	\centering 
	\includegraphics[width=1\linewidth]{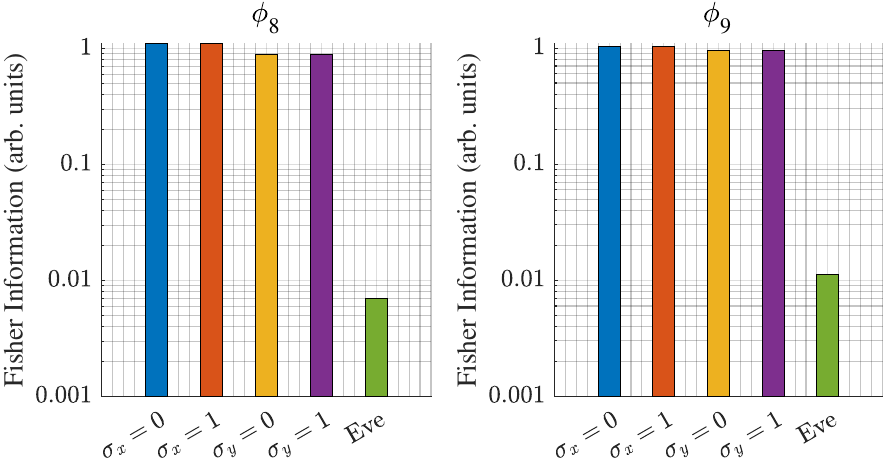}
	\caption{The Fisher information of two selected phases is shown in Fig.~\ref{fig4}. The measured CFI of $\phi_8$ (left) and $\phi_9$ (right) are shown for the four polarization states sent by Alice (labeled as $\sigma_x=0$, $\sigma_x=1$, $\sigma_y=0$, and $\sigma_y=1$). Additionally, the CFI accessible to Eve by eavesdropping on the encrypted data from Bob's side is indicated by the bar labeled ``Eve''. For ideal states, the elicited CFIs for Alice's four states are all equal to 1, while Eve can only obtain a CFI of 0.}\label{fig5}
\end{figure}

To further quantify information leakage, we evaluate the classical Fisher information (CFI), defined as
\begin{multline} \label{eq4}
	{F}_{p} = \frac{1}{p_{i}(out|\phi)} \left( \frac{\partial p_{i}(out|\phi)}{\partial \phi} \right)^2 \\
	+ \frac{1}{1-p_{i}(out|\phi)} \left( \frac{\partial \left[ 1 - p_{i}(out|\phi) \right]}{\partial \phi} \right)^2 \\
	\begin{aligned}
		& \hphantom{\mathcal{F}_{p}} = \frac{\left[\partial p_{i}(out|\phi) / \partial \phi \right]^2}{p_{i}(out|\phi)\left[1-p_{i}(out|\phi) \right]}
	\end{aligned}
\end{multline}
We arbitrarily choose $\phi_8 = 5.515$ and  $\phi_9 = 6.013$ to calculate the CFI (details provided in Appendix~\ref{appendix B}), and the results are shown in Fig.~\ref{fig5}. The analysis reveals significant asymmetry in information gain: on Alice's side, the CFI for each sent polarization state approaches 1, while the information potentially leaked to Eve by accessing Bob's encrypted information is approximately two orders of magnitude lower.

    We also evaluate the precision which can be calculated from the Cram{\'e}r-Rao bound, expressed as $(\delta \phi)^2 \geq 1 / N F_p$, where $F_p$ is the CFI and $N$ is the number of the measurements. For $N \simeq 2.1 \times 10^4$ in our experiment, the total CFI for the four polarization states sent by Alice for $\phi_8$ and $\phi_9$ is approximately 4 for both cases. In contrast, Eve's CFI estimates for $\phi_8$ and $\phi_9$ are 0.008 and 0.011, respectively. Consequently, the precision for $\phi_8$ and $\phi_9$ around 0.0035 rad for Alice, while those for Eve are approximately 0.077 rad and 0.066 rad. These results indicate that the maximum possible information obtainable from Bob’s encrypted data is much less than that available to Alice.

\begin{figure}
	\centering 
	\includegraphics[width=1\linewidth]{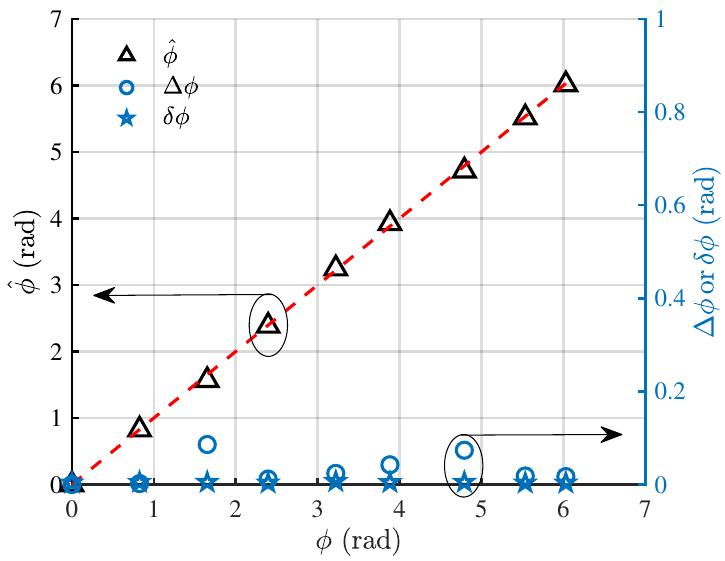}
	\caption{Precision of estimated phases. The black triangles represent the estimated phases. The red dashed line represents where the estimated phase equals the ideal phase. The blue circles represent the deviations between the estimated and ideal phases, calculated as $\Delta \phi=\left|\hat{\phi} - \phi\right|$, and the blue stars denote the Cramér-Rao bound ($\delta \phi$) for each phase.
	}\label{fig6}
\end{figure}

Finally, we estimate the precision of obtained phases. The results are illustrated in Fig.~\ref{fig6}. From the figure, it can be observed that the estimated phases closely match the ideal phases and  the deviations between the estimated and ideal phases, which is calculated using $\Delta \phi=\left|\hat{\phi} - \phi\right|$, is approximate to the Cram{\'e}r-Rao bound.

	\section{Conclusion}\label{conclusion}
In summary,	we have successfully demonstrated an entanglement-free SQRS protocol that can estimate a phase over a long distance of 50 km through optical fiber. This protocol offers practical advantages by utilizing single-qubit states, which are easier to prepare compared to entangled states. To address the challenges posed by practical imperfections, we developed a pre-calibration technique, which enabled the protocol to efficiently estimate the phase in a real-world setup, achieving high precision in phase estimation. Furthermore, we validated the security of the protocol by utilizing the Fisher Information (FI) to quantify the asymmetry in parameter estimation between the legitimate user  and a potential eavesdropper. Our work paves the way for future implementations in various practical scenarios, such as remote health monitoring and secure communication networks.

	\section{Acknowledgments}
	This study was supported by the National Natural Science Foundation of China (Nos. 62171144, and 62031024), the Guangxi Science Foundation (No.2021GXNSFAA220011), and Open Fund of IPOC (BUPT) (No. IPOC2021A02).		

\appendix

\section{CALCULATION OF FISHER INFORMATION}\label{appendix B}		
To validate the security of the protocol, we use the CFI to quantify the asymmetry in parameter estimation between Alice and a third party, Eve. This asymmetry arises because only Alice possesses the specific sequence of the four transmitted quantum states. By exploiting this advantage, Alice can use Bob's measurement results to statistically derive the experimental results summarized in Table~\ref{table1}. In contrast, even if Eve obtains Bob's measurement results through the classical channel, her lack of knowledge about the transmitted states means she can only determine the ratio of response counts between SPD-1 and SPD-2, without discerning their specific distribution characteristics.

\begin{figure}
	\centering 
	\includegraphics[width=0.8\linewidth]{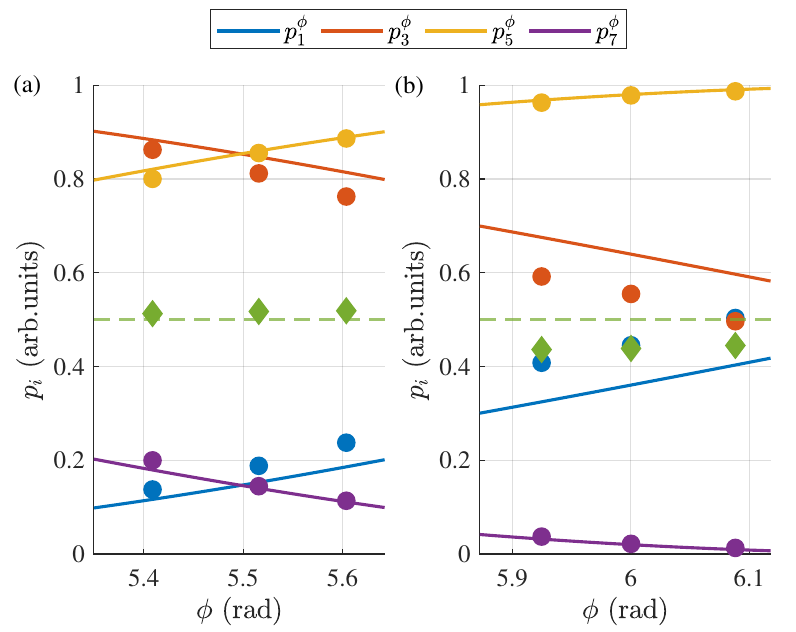}
	\caption{Measurement of CFI at $\phi_8$ (a) and $\phi_9$ (b) in Fig.~\ref{fig5}. Taking $\phi_8$ and $\phi_9$ as the central points, the two neighboring points are used to approximate the derivative at these central points. Thus, the CFI can be calculated from Eq.~(\ref{eq4}).}\label{fig7}
\end{figure}

For a Bernoulli distribution, the CFI can be calculated using Eq.~(\ref{eq4}), where ${F}_{p}$ represents the CFI and $i \in \{1,2,...8\}$. To accurately determine the derivative at the central point, we estimate it by fitting the slope using two nearby points, as shown in Fig.~\ref{fig7}. By substituting the corresponding experimental values into Eq.~(\ref{eq4}), we obtained the CFI for (a) $\phi_8$ and (b) $\phi_9$. Notably, Eve's CFI was calculated using the green diamond markers representing $(n_1 + n_3 + n_5 + n_7) / \sum_{i=1}^8 n_i$ in Fig.~\ref{fig7}.	

\end{document}